\documentclass[12pt,reqno]{article}
\usepackage{amssymb, amsfonts, amsbsy, latexsym, epsfig, color}
\usepackage{amsmath}
\usepackage[ngerman,english]{babel}
\usepackage[authoryear,round]{natbib}

\newcommand{\be}{\begin{equation}}
\newcommand{\ee}{\end{equation}}

\usepackage{soul} 
\newcommand{\del}[1]{\st{#1}}
\newcommand{\intd}[1]{$\lfloor$#1$\rfloor$}

\usepackage{stackengine}
\usepackage{scalerel}
\usepackage{graphicx}
\newlength\lthk
\def\bline{\rule{2ex}{\lthk}}
\def\slash{\rotatebox{60}{\bline}}
\def\parallelogram{\stackMath\scalerel*{%
  \def\stackalignment{l}{\stackunder[-.5\lthk]{%
  \def\stackalignment{r}\stackon[-.5\lthk]{\slash\rule{.866ex}{0ex}\slash}{\bline}}%
  {\bline}}}{\square}%
}

\makeatletter

\newcommand{\zsp}{{\vrule height0ex width0ex depth0pt}}

\newcommand{\@SeitenzaehlungAmRd}[1]{
             \marginpar[\hfill\footnotesize\textnormal{#1}\hspace*{1.5em}]%
                        {\hspace*{1.5em}\footnotesize\textnormal{#1}}%
                        }

\newcommand{\SB}[1]{\@SB{#1}}

\newcommand{\@SB}[1]{%
      \normalmarginpar%
        \zsp\hbox{\hspace{-.15ex}$|$\hspace{-.15ex}}\nobreak
        {\@SeitenzaehlungAmRd{{#1}}}\nobreak
                   }

\newcommand{\SBkeinStrich}[1]{\leavevmode\@SBkeinStrich{#1}}
\newcommand{\@SBkeinStrich}[1]{%
      \normalmarginpar%
        \zsp{\@SeitenzaehlungAmRd{{[#1]}}}\ignorespaces
                   }
\newcommand{\ignore}[1]{} %

\makeatother

\begin{document}

\title{Minkowski's Geometric Explanation of Lorentz Contraction}
\author{Tilman Sauer\thanks{Institute of Mathematics, Johannes Gutenberg University Mainz, D-55099 Mainz, Germany, Email: tsauer@uni-mainz.de}}
\date{Version of August 26, 2026}

\maketitle

\begin{abstract}Hermann Minkowski's original explanation of relativistic length contraction using a space-time diagram and conic section geometry of a standard hyperbola is reconstructed on the basis of unpublished manuscript notes dated to the spring of 1908. The manuscript notes likely reflect Minkowski's preparation of a lecture to physics and mathematics teachers. They reveal Minkowski as an educator as well as a promoter of a mathematization of the physical sciences. I also comment on his interpretation of Einstein's work and on the use of language in his popularization of mathematics.
\end{abstract}

\section{Introduction}

A few months before his unexpected and untimely death, Hermann Minkowski gave a talk to the annual gathering of German scientists and physicians in Cologne. The talk was entitled \emph{Space and Time} and began with the words
\begin{quote}
Gentlemen! The views on space and time which I wish to lay before you have sprung from the soil of experimental physics. Therein lies their strength. Their tendency is radical. Henceforth space by itself, and time by itself, are doomed to fade away into mere shadows, and only a kind of union of the two will preserve an independent reality. \citep{MinkowskiH1909Raum}
\end{quote}
In this talk, Minkowski introduced the now standard notions of a four-dimensional ``world'' whose points are space-time events, he introduced the causal light cone with its distinction of space-like, time-like and light-like vectors, and created a number of pertinent neologisms like world-points, world-lines, world-postulate to a wide audience. 

The opening paragraph of his Cologne lecture has often been cited, and, indeed it gives a vivid image of a new understanding of space and time emerging from the special theory of relativity. It also promoted the use of sophisticated mathematical notions and methods for advanced scientific theorizing. In this respect, the entire lecture and its published text have been called ``a magnificent example of scientific agitprop'' \citep[p.\,59]{WalterS2010World}.

Minkowski's Cologne Lecture (\citeyear{MinkowskiH1909RaumBVerh}) with its subsequent publication \citep{MinkowskiH1909Raum} was a carefully crafted text. In this paper, I want to shed some light on the prehistory of Minkowski's lecture. In doing so I will follow Minkowski's own line of argument and put particular emphasis on his geometric explanation of relativistic length contraction, and I will refer to a number of unpublished manuscripts which up to now have not been investigated in any detail, yet.

These manuscripts are probably the place where Minkowski first explored the use of space-time diagrams for a geometric representation of relativistic space-time features. Galison (\citeyear[p.\,90]{GalisonP1979SpaceTime}) suggested that the paper version of the Cologne diagram was ``the first space-time diagram ever drawn'', \cite{WalterS2008Approach} more specifically suggests that it was the first ``graphical-illustrative technique'' for relativity theory (following a non-space-time diagram by Poincaré). In any case, they reveal a creative process and are of interest for a critical genetic analysis of mathematical texts \citep{HaffnerEEtal2025Atelier}.


Minkowski's Cologne Lecture came about and was intended as a popularization of a more specific research project pursued by Minkowski dealing with a sophisticated mathematical treatment of relativistic electrodynamics. Shortly before his explicit reflections on space and time, Minkowski had completed and published a longer technical paper entitled \emph{The Basic Equations of the Electrodynamics of Moving Bodies} (\emph{Die Grundgleichungen der Elekrodynamik bewegter Körper}) \citep{MinkowskiH1908Grundgleichungen}.

Minkowski's contribution to the early development of relativity theory has been the subject of a number of historical analyses. Early commentaries focussed on a comparison with Einstein's work. Pyenson (\citeyear{PyensonL1977Minkowski,PyensonL1979Physics}) analyzed the local context of Minkowski's work in Göttingen, in particular analyzing a joint seminar on electron theory, conducted together with Hilbert, Wiechert and Herglotz in 1905 and its syllabus. \cite{GalisonP1979SpaceTime} has focussed on the emergence of specific Minkowskian aspects of relativistic space-time. \cite{CorryL1997Minkowski,CorryL2010Minkowski} has emphasized the parallels to Hilbert's ongoing axiomatization project. Walter has analyzed Minkowski's work in a number of studies, placing his work in a larger context of introducing four-dimensional vector methods into physical theory, commenting on the role of Henri Poincar\'e, analyzing Minkowski's attempt to generalize Newtonian gravitation theory, and emphasizing the scandal that Euclidean geometry was no longer to be regarded as sufficient to describe space-time \citep{WalterS1999Minkowski,WalterS1999Style,WalterS2007Breaking,WalterS2008Approach,WalterS2010World,WalterS2018Ether,WalterS2018Figures}.

\section{Prehistory and Context}

In late 1907, when Minkowski was working intensely on his paper on the \emph{Grundgleichungen} and
with his work on relativistic electrodynamics well under way, he also began to think about propagandizing his results. More specifically, the prehistory of his reflections about space and time is spanned by the following events.

\bigskip

He gave a presentation  to the \emph{G\"ottinger Mathematische Gesellschaft} (G\"ottingen Mathematical Society, GMG) on November 5, 1907.\footnote{Apparently, Minkowski had a typescript made of his talk which was produced with several carbon copies. Two versions are preserved in Göttingen (Niedersächsische Staats- und Universitätsbibliothek, Handschriftenabteilung (SUB), Cod.\,Ms.\,Math.Arch.60.3), one of them containing corrections in Minkowski's hand. Sommerfeld later published this lecture on the basis of these copies, selectively integrating Minkowski's handwritten corrections. It appeared posthumously in 1915 in the \emph{Annalen} \citep{MinkowskiH1915Relativitaetsprinzip}. This paper was also reprinted a little later in the \emph{Jahresbericht der Deutschen Mathematiker-Vereinigung} \citep{MinkowskiH1916Relativitaetsprinzip}.} An abstract of his talk in the \emph{Mitteilungen und Nachrichten} of the  \emph{Jahresbericht der Deutschen Mathematiker-Vereinigung} (JDMV-MN) indicates that he introduced an imaginary time coordinate and formulated the electrodynamic field equations in a four-dimensional, Lorentz covariant manner. The abstract also mentions that Minkowski discussed applications of the relativity principle to the theory of gravitation at the end of his presentation.
However, this abstract does not yet contain a hint about the possibility of explaining Lorentz-contraction of electrons by means of conic section geometry of a hyperbola. Minkowski introduced a four-dimensional viewpoint, imaginary time, and presumably also matrix notation \`a la Cayley, as was done in the long paper.

A second lecture to the GMG followed quickly thereafter, on December 10. This one was coordinated with a lecture by Felix Klein on quaternions and biquaternions. Another abstract in the JDMV-MN summarized this talk. It reports that he unified the electromagnetic field equations into a single biquaternionic equation (``eine einzige Biquaternionengleichung''). Again, Minkowski appears to have taken pride in the application of advanced and sophisticated mathematical concepts to problems of theoretical physics. However, even though Minkowski must have been in the midst of writing up the manuscript of his long article, its published version contains no reference to quaternions or biquaternions.

Then on 21 December 1907, Minkowski presented his paper to the \emph{Göttinger Gesellschaft der Wissenschaften} (Göttingen Society of Sciences or `Academy' for short) for publication in its \emph{Nachrichten}. The final manuscript of his paper was delivered to the typesetter only some time later but the formal presentation of his manuscript to the Academy for publication gave Minkowski a first occasion for pointing out the relevance of his work to a broader audience. A manuscript and a typescript of his brief remarks presenting his paper to his Academy colleagues are preserved in the Archives \citep{SauerT2026Promoting}.
In this brief presentation, Minkowski established a certain historiography of this field of mathematical physics, placing his own work into a line starting with Hertz's version of electrodynamic field equations of 1890, over Lorentz's electrodynamics, and Einstein's 1905 work on the relativity principle. He emphasized that mathematicians had created concepts ahead of time which were now being employed in understanding modern developments in theoretical physics. The suitability of the mathematical concepts and framework is described with a reference to Leibniz as `preestablished harmony.' Again, in his presentation to the Academy there is not yet a hint about the geometric explanation of the Lorentz contraction.

Such was the situation at the end of the year 1907. Minkowski had presented his manuscript to his Academy colleagues for publication in the Academy's proceedings. He was still editing his final manuscript. He was also possibly thinking ahead to follow up on his first paper by a second technical paper discussing the electrodynamics of moving media. In early 1908, Minkowski continued working on space and time and relativistic electrodynamics and 
for the following, it will be useful to point at some further relevant developments and events. 

Minkowski knew about and cited Einstein's famous paper from 1905 on the \emph{Electrodynamics of Moving Bodies} \citep{EinsteinA1905Elektrodynamik}, which introduced the special theory of relativity. The paper had appeared too late to be included in the syllabus of the joint seminar on electron theory which Minkowski had conducted together with David Hilbert, Emil Wiechert and Gustav Herglotz in 1905 \citep{PyensonL1979Physics}. But on October 7, 1907, Minkowski had written to Einstein, asking him for offprints of this paper \citep[Doc.\,62]{CPAE05}. In the meantime, Einstein himself had been working on a long review of his theory of (special) relativity \citep{EinsteinA1907Relativitaetsprinzip}. His own review had been received by Johannes Stark,  editor of the \emph{Jahrbuch f\"ur Radioaktivit\"at und Elektronik}, on December 4, 1907, and the published paper had come out on January 22, 1908 \citep[p.\,485]{CPAE02}. We don't know when Minkowski learned about Einstein's \emph{Jahrbuch} review, but we may surely assume that Minkowski would have studied Einstein's review as soon as he could get a hold of it.  

Another paper, that Minkowski would have studied is Max Planck's \emph{On the Dynamics of Moving Systems} \citep{PlanckM1907Dynamik}. Planck had already published it in the \emph{Proceedings} of the Berlin Academy in 1907. But on March 6, 1908, Planck also submitted his same paper to be reprinted in the \emph{Annalen der Physik} \citep{PlanckM1908Dynamik}.

On February 21, Minkowski delivered the final manuscript of the \emph{Grundgleichungen} to the printer \citep[p.\,219]{WalterS2007Breaking}.  Then, on April 5, the G\"ottingen Academy's \emph{Nachrichten} containing his \emph{Grundgleichungen} paper was issued.\footnote{See the page following p.\,128 in the relevant year volume of the \emph{Nachrichten}.} A few days later, Minkowski travelled to Rome to participate in the IVth International Congress of Mathematicians. 

\section{The manuscript}

Sometime between the becoming available of Einstein's \emph{Jahrbuch} paper and the publication of Minkowski's \emph{Grundgleichungen} paper, i.e.\ between January 22 and April 5, Minkowski wrote a manuscript, in which he explained, for the first time, the Lorentz contraction using a space-time diagram.\footnote{SUB, Doc.\,Ms.\,Math.\,Arch.\,60.4, f50--65. The pages are sequentially numbered by Minkowski in the upper right corner with simple numbers for pages 1 through 10, and with numbers enclosed by dashes (- nn- ) for pages 11 through 16. In citing from these manuscript pages, I will use a dual reference fxx/Myy, indicating both the folio number \emph{xx}, found as pencil notes in the right margin of the recto pages, and (when it exists) the page numbering \emph{yy} that Minkowski himself put in the right top corner.}  An  approximate date for that manuscript can be gathered by two remarks. On f51/M2, Minkowski wrote:
\begin{quote}
As Einstein continues to work on his chosen path, he recently published a long paper in Stark's \emph{Jahrbuch der Radioaktivit\"at und Elektronik}: ``On the Relativity Principle and the Conclusions Drawn from It'', in which he now strives for a further extension of his methods.\footnote{%
``Indem Einstein in seiner eingeschlagenen Bahn weiter arbeitet, es ist jüngst von ihm in dem Starkschen Jahrbuch der Radioaktivität und Elektronik ein längeres Referat erschienen: Ueber das Relat.\ und die aus demselben gezogenen Folgerungen, in dem er nun eine weitere Ausdehnung seiner Methoden anstrebt.'' The reference is to \citep{EinsteinA1907Relativitaetsprinzip}.}
\end{quote}
This reference clearly establishes the publication of Einstein's (\citeyear{EinsteinA1907Relativitaetsprinzip}) paper, January 22, 1908, as a terminus post quem, and on f54/M5, Minkowski wrote:
\begin{quote}
In a longer paper, which will appear in the \emph{Göttinger Nachrichten} these days, I show in particular how this principle of relativity leads to decisive results about electrodynamics of moving bodies.\footnote{%
``In einer längeren Arbeit, die in diesen Tagen in den Gött. Nachr. erscheint, insbesondere zeige ich, wie dieses Relativitätsprinzip insbesondere zu entscheidenden Ergebnissen über Elektrodynamik bewegter Körper führt.''}
\end{quote}
This reference puts his manuscript close to the issuing of his own paper on April 5 as a terminus ante quem. Also at the end of the manuscript, he has
\begin{quote}
For bodies at rest there is agreement, for moving bodies there is disagreement. I now believe that my work, which will be published in the next few days, will bring a definitive decision. The principle of relativity, as I show in it, leads to very specific equations that also deviate from Lorentz's equations.\footnote{%
``Für ruhende Körper herrscht Übereinstimmung, für bewegte Meinungsverschiedenheiten. Ich glaube nun, dass meine Arbeit, die in diesen Tagen erscheint, definitive Entscheidung bringt. Das Relativitätsprinzip, wie ich darin zeige, führt auf ganz bestimmte Gleichungen, die auch von den Lorentzschen abweichen.'' (f65/M16)}
\end{quote}

This manuscript, I contend, gives the first derivation of Lorentz contraction by a geometric argument using conic section geometry of a hyperbola. It also documents the evolution of Minkowski's space-time diagrams.

The manuscript clearly was written as notes for a presentation. It begins with the usual abbreviation ``M.H.'' for ``Meine Herren'' or ``Gentlemen'' and continues like this (see Fig.\,\ref{fig:f50M1}):
\begin{figure}
\begin{center}
\includegraphics[width=0.9\linewidth]{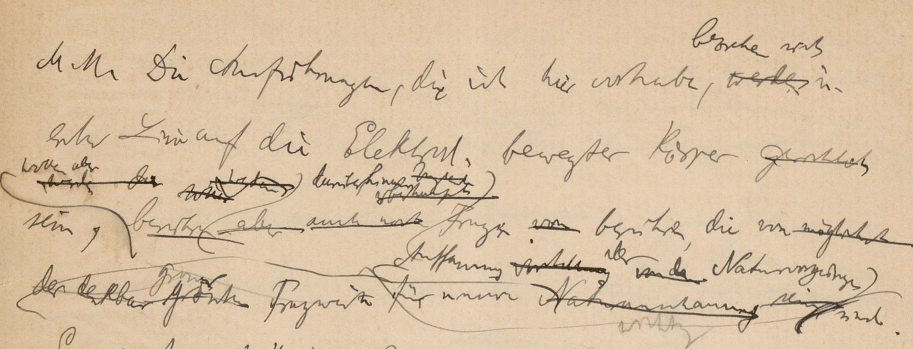}
\caption{The beginning of Minkowski's manuscript from late March\,/\,early April  1908 (SUB Cod.\,Ms.\,Math.\,Arch.\,60.4, f50, M1).}
\label{fig:f50M1}
\end{center}
\end{figure}
\begin{quote}
M.H. The remarks I intend to make here are primarily concerned with the electrodynamics of moving bodies, but I also want to touch on questions that are of great importance for our understanding of natural processes in general.\footnote{A more diplomatic transcription of the first sentence would be:\\
``M.H. Die Ausf\"uhrungen, die ich hier vorhabe, \del{werden} \intd{beziehen sich} in erster Linie auf die Elektrod. bewegter K\"orper \del{gerichtet} sein, \del{berühren aber} \intd{\del{werden dar} \intd{wollen aber \del{sollen} \del{werden} darüberhinaus \del{zugleich} \del{überhaupt}}}  auch noch Fragen \del{von} berühren, die von \del{möglichst} \del{der denkbar grössten} \intd{grosser} Tragweite für unsere \del{Naturanschauung} \intd{Auffassung \del{Vorstellung} \del{von den} \intd{der} Naturvorgänge} \intd{wichtig} \del{sein} sind.'' (f50/M01)\\
In all quotes from the manuscripts I give what I take to be the intended final version without indicating deletions or interlineations, unless they are particularly significant or interesting.}
\end{quote}
There is no hint in the manuscript as to what the occasion or audience of this talk would have been.
The printed titles and abstracts of talks presented to the GMG do not contain any indication of this talk or of any talk by Minkowski in the relevant time period. It might have been presented at an adhoc meeting of the GMG, or at some other gathering of G\"ottingen mathematicians. Very likely it may, however, be a first draft for a lecture he gave at a \emph{Ferienkurs} (vacation course).\footnote{Pyenson (\citeyear[p.\,72]{PyensonL1977Minkowski}) citing a passage about Einstein from p.(3) of the manuscript claims, without further justification, that it was ``during a series of lectures on the electrodynamics of moving bodies delivered before an audience of physicists in the spring of 1908''. Walter (\citeyear[p.\,47]{WalterS1999Minkowski}) also suggested that the pages around f52 were notes for these Ferienkurse. However, what I take to be the actual notes for these Ferienkurse are conjectured by him to be first drafts of the Cologne lecture \citep[p.\,49]{WalterS1999Minkowski}.}  Such courses were announced to take place in Göttingen from 21 April to 2 May, see JDMV-MN (1908, pp.11--12). They were intended for secondary school teachers (``Lehrer höherer Schulen'') and the program included a number of lectures on mathematics and physics as well as visits and demonstrations of various places like the mathematics library (``Mathematisches Lesezimmer''), the collection of mathematical models, the observatory, and various physics laboratories. In the program, Minkowski was announced to give a two-part lecture (``2 Doppelst.'') on ``Recent Ideas on the Fundamental Laws of Mechanics'' (``Neuere Ideen über die Grundgesetze der Mechanik.''). The Minkowski papers also contain a manuscript for that lecture.\footnote{That second manuscript, as found in the archival folder SUB Cod.Ms.\,Math.Arch.\,60.4, begins right after the end of the other manuscript. However, its subsequent pages are dispersed over two folders (60.4 and 60.2) and are no longer preserved in their original sequence. My reconstruction of this manuscript is the following: First part: 60.4: f1, 60.2: f1--f6, 60.4: f7, f67--f68; these pages are numbered consecutively from 1 to 10 by Minkowski. After 60.4, f68 (p.\,10) there are several blank sheets, and I have not been able to locate any page that would follow after this one. Second part: 60.4: f19--f20, 60.2: f7--f8, 60.4: f6, f21--f23, f75--f76, f18; again these pages are numbered consecutively from 1 to 12, however, p.\,3 is missing and could not be located. } They indicate a date of 24 and 25 April (f66) and are also clearly addressed to physics teachers.\footnote{Minkowski also mentioned those vacation courses in a letter to Hurwitz, dated 10 May, 1908: ``auch hatte ich noch einen Ferienkurs abzuhalten.'' SUB Cod.\,Ms.\,Math.Arch.78:212. In later notes for his lecture to the GMG of July 28, 1908, he appears to refer to the \emph{Ferienkurs} lecture as a lecture at the ``Physical Society'' (Physikalische Gesellschaft'') Math.Arch.60.4, f27, f39.} In any case, between these two manuscripts, Minkowski travelled to Rome for the International Congress of Mathematicians, where he met, among others, H.A.\,Lorentz and H.\,Poincaré.\footnote{The Congress took place from  6--11 April 1908; see \cite[p.\,238]{KoxAJ2008Correspondence} and p.\,263 for Lorentz on Minkowski's death. The meeting between Lorentz and Minkowski is mentioned in Math.Arch.60.4, f11, see also \citep[p.\,67]{WalterS1999Minkowski}, who establishes that Minkowski and Lorentz met at the occasion.}

Regardless of whether the first manuscript records notes for a talk of its own or was a first draft for the \emph{Ferienkurs} lecture, it seems likely that Minkowski, studying Einstein's \emph{Jahrbuch} paper, was realizing the possibility of a purely geometric explanation of relativistic length contraction. There are indications that Minkowski himself might have been rather excited about his new insight. He wrote:
\begin{quote}
To state my own opinion right away, I contend that Lorentz's hypothesis, correctly understood, represents a law of nature of the very first rank, I would even like to call it the first of all laws of nature; this because it concerns the most basic concepts of all our knowledge about nature, our conception of space and time, and also because of quite extraordinary consequences of this law, the most astonishing of which have not even been discovered yet.\footnote{%
``Nämlich gleich meine eigene Meinung zu sagen, so bedeutet jene Lorentzsche Hypothese richtig aufgefasst, ein Naturgesetz allerersten Ranges, ich möchte sogar geradezu sagen, das erste aller Naturgesetze, nämlich einerseits deshalb, weil es sich um die ursprünglichsten Begriffe aller Naturerkenntnis, um die Auffassung von Raum und Zeit handelt, dann wegen ganz ausserordentlicher Konsequenzen dieses Gesetzes, von denen die überraschendsten bisher noch gar nicht bemerkt worden sind.'' (f52/M3)}
\end{quote}

\subsection{Minkowski's analysis of Lorentz contraction}

If this was indeed the first time that Minkowski presented---or intended to present---to a public audience his geometric derivation of Lorentz contraction, it will be worthwhile to analyze this derivation in somewhat more detail. I want to emphasize in the following, going along Minkowski's notes, that this crucial insight actually depends on a number of assumptions and abstractions that are involved in the argument, and I will analyze the argument in some detail.

\paragraph{Identification of a problem.} The first step is to recognize a significant problem that offers itself to treatment by sophisticated mathematical methods. Minkowski begins his presentation with a discussion of Lorentz's hypothesis that moving bodies appear contracted in the direction of motion if viewed by an observer at rest. He mentions Einstein's work on the relativity principle as a continuation of Lorentz's work. The relativity principle, in particularly, emphasizes that the surprising contraction hypothesis does not depend on a  distinction between a special system at rest and a moving system. Rather the same effect is valid whatever system is taken to be at rest, and any uniformly moving system can be taken as a rest system.

This surprising hypothesis of a physical contraction of rapidly moving bodies is associated with a mathematical feature. The relevant foundational electrodynamical equations are of such a character that they remain invariant under a certain group of coordinate transformations, and these transformations are characterized by the fact that they leave the expression
\begin{equation}
x^2+y^2+z^2-c^2t^2
\label{eq:xyzt}
\end{equation}
invariant. Here $x$, $y$, $z$ are spatial coordinates (now called space-like using the term introduced in the Cologne lecture), $t$ is a time coordinate (later time-like), and $c$ is the velocity of light in vacuum. In his manuscript, Minkowski does not explicitly point out the invariance of this expression. Although the precise knowledge and Minkowski's understanding of the early relativity literature is a topic of debate,
we may as a first approximation assume that he was well acquainted with most of the relevant literature on electrodynamic theory, especially electron theory, by Max Abraham, Lorentz, Poincar\'e and others.\footnote{See Pyenson's (\citeyear{PyensonL1979Physics}) analysis of the syllabus for the 1905 seminar, 
see also \citep{WalterS2008Approach}.}

In the manuscript, he announced:
\begin{quote}
My first task today is to explain to you in what sense Lorentz's contraction hypothesis means nothing other than a new conception of time.\footnote{%
``Ich stelle mir nun als erste Aufgabe, Ihnen heute auseinanderzusetzen, in welchem Sinne die Lorentzsche Kontraktionshypothese nichts anderes als eine neue Auffassung des Zeitbegriffs bedeutet.'' (f52/M3)}
\end{quote}
And in an afterthought, he added in the margin:
\begin{quote}
I could start from Lorentz's postulate and then derive the principle of relativity. However, it is probably easier to understand the opposite approach, in which I first formulate the principle of relativity and then show you how Lorentz's contraction assumption is derived from it.\footnote{%
``Ich könnte da von der Lorentzschen Forderung ausgehen und daraus auf das Relativitätsprinzip kommen. Leichter aufzufassen ist aber wohl der umgekehrte Weg, dass ich Ihnen zuerst das Relativitätsprinzip formuliere und hernach zeige, wie sich daraus die Lorentzsche Kontraktionsannahme ableitet.'' (f51v)}
\end{quote}

After identifying an explanandum, one can now proceed to develop the argument. Minkowski announced his plan:
\begin{quote}
I will now try to bring you to an understanding of the principle of relativity by presenting more details. I will make use of simple figures which, however, will require a new kind of abstraction.\footnote{%
``Ich will nun zunächst Sie zum Verständnis des Relativitätsprinzipes durch genauere Details zu führen suchen. Ich werde mich einfacher Figuren bedienen, die aber schon eine neuartige Abstraktion erheischen werden.'' (f54/M5--f55/M6) }
\end{quote}

\paragraph{Reduction of spatial dimensions.} The first such step of abstraction is a dimensional reduction. In order to be able to draw meaningful figures---we would now say: space-time diagrams---, we have to reduce the number of relevant spatial dimensions:
\begin{quote}
In order to be able to draw anything at all, I have to assume very simple conditions, i.e.\ I will conceive of the world as one-dimensional instead of three-dimensional.\footnote{%
``Um überhaupt Zeichnungen entwerfen zu können, muss ich die Verhältnisse sehr einfach annehmen und zwar werde ich mir die Welt statt dreidimensional nur eindimensional denken.'' (f55/M6)}
\end{quote}
To make it more concrete, Minkowski imagined one-dimensional electrons, and introduced a term ``worldline'' (``Weltgerade'') which here refers to this one-dimensional manifold:
\begin{quote}
In space, everything should take place on a straight line, the electrons are bodies that change their position on the worldline. Here, the black one, is an electron at rest, the red one a moving one, contracted according to Lorentz, the green one an even faster moving one.\footnote{%
``Räumlich soll sich alles auf einer Geraden abspielen, die Elektronen seien Körper, die auf der Weltgeraden ihren Ort verändern. Hier [das schwarze] ein ruhendes Elektron, das rote ein bewegtes, nach Lorentz kontrahiertes, das grüne ein noch schneller bewegtes.'' (f55/M6)}
\end{quote}
Obviously, Minkowski was intending to use colored chalk for drawing pictures at the black board during his presentation. His manuscript, though, was written with black ink and he only indicated a little sketch in the margin:

\begin{center}
\includegraphics[width=0.4\linewidth]{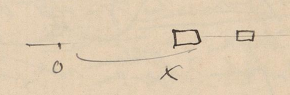}
\end{center}

\noindent
Using the \emph{Weltgerade} as one Cartesian coordinate axis by singling out an origin on it, positions of the electron are given by their $x$ coordinate. Note that the term ``Weltgerade'' is distinct from the term ``Weltlinie,'' which he later introduced in his Cologne lecture.

\paragraph{Introduce time as a coordinate axis.} The next step in creating space-time diagrams is to exploit the two-dimensionality of the paper plane and introduce a second Cartesian coordinate representing time:
\begin{quote}
Let us represent the movement of an electron graphically by plotting not only the locations but also the times. I will therefore mark the time on a second axis. I emphasize that these figures now represent a certain abstraction.\footnote{%
``Suchen wir uns die Bewegung eines Elektrons graphisch darzustellen, indem wir nicht bloss die Orte auch die Zeiten eintragen. Auf einer zweiten Axe werde ich also die Zeit markieren. Ich betone, dass diese Figuren jetzt eine gewisse Abstraktion bedeuten.'' (f55/M6)}
\end{quote}
The existence of material points or bodies in time, moving or not moving, then leads to their representation as continuous lines:
\begin{quote}
So now every electron exists in space and time. We now enter locations at the different times, so this line  e.g.\ where we have the same $x$ at all $t$ becomes an electron at rest, this one a uniformly moving one, this one a non-uniformly moving one.\footnote{%
``Also es existiert nun jedes Elektron in Raum und Zeit. Wir tragen nun Orte in den verschiedenen Zeiten ein, so wird diese Linie z.B.\ wo wir dasselbe $x$ zu allen $t$ haben ein ruhendes Elektron, dieses ein gleichförmig vorstellen, \del{dieses ein ungleichförmig bewegtes}.'' (f56/M7)}
\end{quote}
In the manuscript, the reference to non-uniformly moving electrons was deleted, presumably because of the conceptual problem of defining the geometric representation of an electron that is not moving with constant velocity along a straight line (see the discussion below). But in the margin, Minkowski again sketched an illustration of the situation for the black board:

\begin{center}
\includegraphics[width=0.3\linewidth]{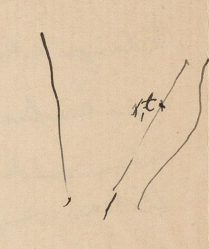}
\end{center}

It is at this point that he introduced the notion of what he would later, in his Cologne talk, famously call a world line. Here he calls it simply a space-time line:
\begin{quote}
I call such a line, that corresponds to the entire existence of a point-like body at all times, a space-time line.\footnote{%
``Eine solche Linie, wie sie die der ganzen Existenz eines punktförmigen Körpers zu allen Zeiten entspricht, nenne ich eine Raum-Zeitlinie.'' (f56/M7)}
\end{quote}
The manuscript at this point shows some heavy revisions, apparently because Minkowski deliberated whether he should, already at this point, emphasize the arbitrariness of coordinates in contrast to the reality of such space-time lines of material bodies.

\paragraph{Introduce a suitable unit of time.} The next step in the preparation of the argument involves the introduction of a unit of time, such that the resulting space-time diagrams may distinctly express relativistic effects. The now familiar light cone is obtained by setting the velocity of light equal or close to 1, or, in Minkowski's words:
\begin{quote}
Now we want to define the unit of time such that the speed of light is 1, i.e.\  as $\frac{1}{3\cdot 10^{10}}$ sec with 1cm as unit of length. This appears at first a very inconvenient unit of time, but now everything is much simpler in the graphical forms.\footnote{%
``Nun wollen wir vor Allem die Zeiteinheit so festlegen, dass die Lichtgeschwindigkeit 1 ist, also bei 1cm als Längeneinheit als $\frac{1}{3\cdot 10^{10}}$ sek. das ist ja zunächst eine sehr unbequeme Zeiteinheit, aber nun verläuft alles in den Formen sehr viel einfacher.'' (f56/M7)}
\end{quote}
For the geometric argument, it is not necessary to introduce an imaginary time coordinate and he does not mention this in his first manuscript. But Minkowski had introduced an imaginary time in his 1908 paper, and he also mentioned it in his second manuscript. There he says:
\begin{quote}
One could think that in the laws of nature the time $\frac{1}{3\cdot 10^{10}}$ sec plays the same role as $\sqrt{-1}$cm.\footnote{Man \del{kann} \intd{könnte denken} sagen, es spielt in den Naturgesetzen die Zeit $\frac{1}{3\cdot 10^{10}}$ sec die [Rolle] wie $\sqrt{-1}$cm. (f7/M8) Similarly, he wrote to his friend Hurwitz on February 1, 1908: ``Wenn man $\frac{1}{3\cdot10^{10}}$\,sekunde als $\sqrt{-1}$\,centimeter bezeichnet, [d.h.\ die Zeiteinheit so gewählt, dass die Lichtgeschwindigkeit 1 wird, und sie dann als $\sqrt{-1}\times$Längeneinheit anspricht], so wird die ganze Physik eine mathematisch höchst befriedigende Wissenschaft.'' (SUB Cod.\,Ms.\,Math.-Arch.78:211)}
\end{quote}

\begin{figure}
\begin{center}
\includegraphics[width=0.7\linewidth]{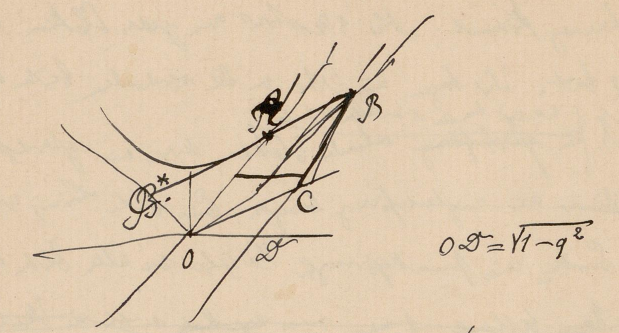}

\includegraphics[width=0.3\linewidth]{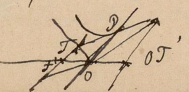}
\includegraphics[width=0.4\linewidth]{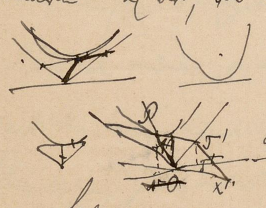}

\caption{Several versions of space-time diagrams with standard hyperbola on f56/M7 and f57/M8.}
\label{fig:hyperbola}
\end{center}
\end{figure}

\paragraph{Introduce standard hyperbola.} The next crucial step is to interpret the invariance of expression (\ref{eq:xyzt}) in terms of the space-time diagram. Since the $y$- and $z$-coordinates have been neglected and since $c$ has been set to unity, $c=1$, the expression defines a hyperbola if set equal to a constant, say -1, such that one gets
\begin{equation}
t^2-x^2=1.
\label{eq:hyperbola}
\end{equation}
In the space-time diagram this equation represents a vertical, equilateral hyperbola. Minkowski drew only the upper branch ($t>0$) of it. There are, in fact, various drawings on his manuscript notes at this point, see Fig.\,\ref{fig:hyperbola}.

The introduction of the standard hyperbola is equivalent to the introduction of a Lorentz signature metric but much of the argument remains euclidean and is only using properties of conic section geometry. In his second manuscript, Minkowski adresses this point, addressing his audience of teaching professionals in a curious way:
\begin{quote}
You will see that the mathematical argument essentially is using properties of a hyperbola, \del{and since you, gentlemen, are wont to question anything that comes to your attention}, such that most of what I am going to say could on occasion be used for embellishing a chapter of analytical geometry, if only some of the conceptual difficulties are downplayed a bit.\footnote{%
``Sie werden sehen, dass die mathematischen Ausführungen wesentlich auf Eigenschaften einer Hyperbel hinauskommen werden, \del{und da Sie, meine Herren gewohnt sind, bei allem, was Ihnen zu Ohren kommt, sich zu fragen}, sodass sogar das Meiste, was ich zu sagen haben werde, wenn man einige damit verbundenen begrifflichen Schwierigkeiten nicht zu stark eintaxiert, sich gelegentlich zur Ausschmückung eines Kapitels der analytischen Geometrie vorbringen liesse.'' (60.2, f5).}
\end{quote}

\paragraph{The axiom of a limit velocity.} At this point, Minkowski introduced an explicit axiom. In fact, it is the only axiom that he refers to as such explicitly:
\begin{quote}
A first axiom is now that nowhere in the world is there a speed greater than the speed of light; the speed of light itself is only an ideal limiting case.\footnote{%
``Ein allererstes Axiom ist nun, dass nirgends in der Welt eine Geschwindigkeit existiert, die grösser als die Lichtgeschwindigkeit ist, die Lichtgeschwindigkeit selbst ist nur ein idealer Grenzfall.'' (f56/M7--f57/M8)} 
\end{quote}
Note that Minkowski here argues in a different way than Einstein did and than would become standard later. He did not identify the principle of relativity and the principle of the constancy of the speed of light as the two defining principles for the special theory of relativity. Instead, he singled out Lorentz's contraction hypothesis as a fundamental postulate, and the existence of a limiting velocity as an axiom.

In his second manuscript, he does not refer to it as an axiom but as a ``first fact'', which here would be ``most important'' and the assumption of which ``perhaps you might be most reluctant to accept at first.''\footnote{``gegen deren Annahme man zunächst sich am meisten sträuben mag.'' (60.2, f7)} But with the acceptance of this limit velocity, Minkowski emphasized, most of the work has been done.\footnote{``Nun ist das wesentlichste mit der Einführung dieser Geschwindigkeitsschranke geschehen.'' (60.4, f8).}

The axiom can be interpreted in the space-time diagram by means of the hyperbola (\ref{eq:hyperbola}). Any space-time line of a material point is of such a nature that its derivative at any point creates a tangent steeper than the asymptotes of the hyperbola, $\frac{dt}{dx}>1$, or $\frac{dx}{dt}<1$. Its direction is given by the vector pointing from the origin to the relevant point on the hyperbola.

\paragraph{Interpret conjugate diameters as coordinate axes.} It is already at this point that Minkowski can bring his graphical representation to bear on the argument. From the theory of conic sections, it is known that the hyperbola is a graph of the equation (\ref{eq:hyperbola}) but $x$ and $t$ need not be conceived of necessarily as the orthogonal Cartesian coordinates one may have started out with. In fact, the coordination may refer to any set of conjugate diameters of the hyperbola. These can be geometrically constructed for any conic section and for a hyperbola will be pairs of diameters symmetric with respect to its asymptotes. On the other hand, space-time points are real and do not depend on the choice of coordinate generating diameters. Any line from the origin through a point on the hyperbola can, on the one hand, represent the space-time line of a uniformly moving material point, and, on the other hand and at the same time, one of the conjugated diameters for the hyperbola.

Conjugate diameters can be constructed geometrically using its asymptotes. These latter correspond to the space-time lines of light and represent the limiting velocity. Conjugate diameters are now constructed for lines connecting the origin and any point on the standard hyperbola by completing a parallelogram with a tangent line at the hyperbola as another side and the asymptote as one of its diagonals.

Minkowski emphasized this point as a feature which would be natural for mathematicians but perhaps difficult to realize for those not trained in mathematics. In his notes, he draws an interesting analogy to the case of identifying diameters for a sphere for beings who would live in a limited neighborhood of one point on the sphere and for whom a certain diameter of the sphere would therefore be the natural one.

\paragraph{Introduce electrons as extended objects.} So far, we have been dealing with point-like objects and their representation in a diagram of a 1+1-dimensional space-time. A next step in the argument is to introduce the notion of extended objects. An extended material object corresponds to a finite interval on the \emph{Weltgeraden}. Any one of the points in that interval creates its own space-time line, and Minkowski calls the set of space-time lines belonging to an electron of finite spatial extension a space-time thread (``Raum-Zeit-Faden''):
\begin{quote}
Let us now see how our new concept of place and time can replace Lorentz's contraction hypothesis. We now have to say what an electron is. Well, I have already spoken of a space-time line. If we think of a one-dimensionally extended body and the space-time line for each material point of it, I will call the totality of the relevant lines a space-time thread.\footnote{%
``Sehen wir nun zu, inwiefern unsere neue Auffassung von Ort und Zeit die Kontraktionshypothese von Lorentz ersetzt. Da haben wir nun zu sagen, was ist ein Elektron. Nun ich sprach schon von einer Raum-Zeitlinie. Denken wir uns einen eindimensional ausgedehnten Körper und für jeden materiellen Punkte desselben die Raum-Zeitlinie, so will ich die Gesamtheit der betreffenden Linien einen Raum-Zeitfaden nennen.'' (f58/M9--f59/M10)} 
\end{quote}
The notion of a space-time thread here reminds of the modern parlance (for more than one spatial dimension) of a world tube. But there was a problem here for Minkowski, which he needed to address before going on. He was certainly familiar with contemporary debates about the shape of a moving electron. Various options had been proposed: a fully rigid electron, one that contracts only along its direction of motion, or one that deforms in a volume-preserving way. Therefore, he had to be more precise about the definition of an electron according to Lorentz as it pertains to the length contraction hypothesis. He needed to translate the notion of a rigid electron into the geometric framework. He also observed that the notion of a Lorentz electron really had only been defined at this point for uniform velocities.  A discussion of extended bodies in the geometric framework therefore involved the explication of the notions of size and of units of length.

In order to address the issue of the proper definition of a Lorentz electron, Minkowski now defined the notion of a normal cross section and he then defined a Lorentz electron for arbitrary motion as one that has a constant normal cross section:
\begin{quote}
the definition would be: it is a space-time thread in which the `normal' cross-section is constant, which should mean the following: If we are at any point of the space-time thread, we draw a line parallel to the course of the space-time thread, and to this direction [we draw] the direction conjugate to the basic hyperbola; the width $QQ'$ of the thread in this direction, but measured by the parallel length $AB$ there of the hyperbola as a unit, should be called the normal cross-section at that point.\footnote{%
``die Definition würde sein: es ist ein Raum-Zeitfaden, bei welchem der \glqq{}normale\grqq{} Querschnitt konstant ist, d.\ soll folgendes bedeuten: Sind wir an irgendeiner Stelle des Raum-Zeitfadens, so ziehen wir  eine Linie parallel dem Verlaufe Raum-Zeitfadens, und zu dieser Richtung die an der Grundhyperbel konjugierte Richtung, die Breite $QQ'$ des Fadens in dieser Richtung, aber gemessen in der parallelen Länge $AB$ die der Hyperbel als Einheit, soll der \glqq{}normale\grqq{} [Normal-]Querschnitt an der Stelle heissen.'' (f60/M11)} 
\end{quote}
In other words, we take the tangent of the space-time line at any point of the space-time thread, look at a line parallel to it that passes through the origin, and then find the point where this line intersects the hyperbola. We take this line as one of the diameters and measure the width of the thread along its conjugate diameter, i.e.\ we take a line parallel to the tangent at the hyperbola at that point of intersection which passes through the point of the space-time thread. The size of the space-time thread is then to be measured along this conjugated coordinate.

While we now are in a position to say what it means that the normal cross section of a space-time thread is constant, we still need to define units of length along our coordinate axes. For a unit hyperbola, these are given by the parallelograms going through the origin and the pertinent point on the hyperbola, and tangents to the hyperbola. Unit distances are then the distance from the origin to the hyperbola for the time axis, and from the hyperbola to the asymptotes along the tangents at the hyperbola for the space axis.

\paragraph{Take into account change of units for different coordinate axes.}  Now all relevant conceptual and graphical elements have been introduced to discuss Lorentz contraction. For this purpose, one draws the space-time threads of two electrons in relative uniform motion. The graphical representations of these threads are extended strips of parallel lines which are separated by a certain distance determined by the size of the electrons. One can now identify cross sections of those strips by lines parallel to the spatial coordinate axis of either one or the other electron. The length of these cross sections can  be measured by unit lengths associated with either one of the spatial axes. A length contraction is then expressed by the geometric fact that cross sections of the same electron have different measures, and more precisely, the size of the cross section along the spatial axis associated with the electron itself is always larger than the size associated with cross section associated with the moving electron. In other words, the electron appears smaller for a moving observer.

\paragraph{Derive the correct numerical contraction factor.}
The final task now is to geometrically deduce the length ratio of the two corresponding cross sections and establish that it is numerically equal to $\sqrt{1-q^2}$ regardless which electron is taken to be at rest and which one is taken to be in relative uniform motion $q$.

This part was not given explicitly by Minkowski in his notes nor in the published version. Sommerfeld later felt that a more explicit argument was needed and explained it in his notes to a reprint of Minkowski's Space and Time paper in the collection \emph{Das Relativit\"atsprinzip}.\footnote{See also \citep[p.\,91]{GalisonP1979SpaceTime} for comments on Sommerfeld's explanation.} Clearly, the argument presupposes a familiarity with conic section geometry that is no longer a given for modern readers, and Minkowski's argument in his oral presentation may have been rather brief and elegant. In the following, I give a detailed putative reconstruction of a possible argument taking clue's from Sommerfeld's notes.

\begin{figure}
\begin{center}
\includegraphics[width=0.55\linewidth]{figures/f56M7vfig1.pdf}
\includegraphics[width=0.40\linewidth]{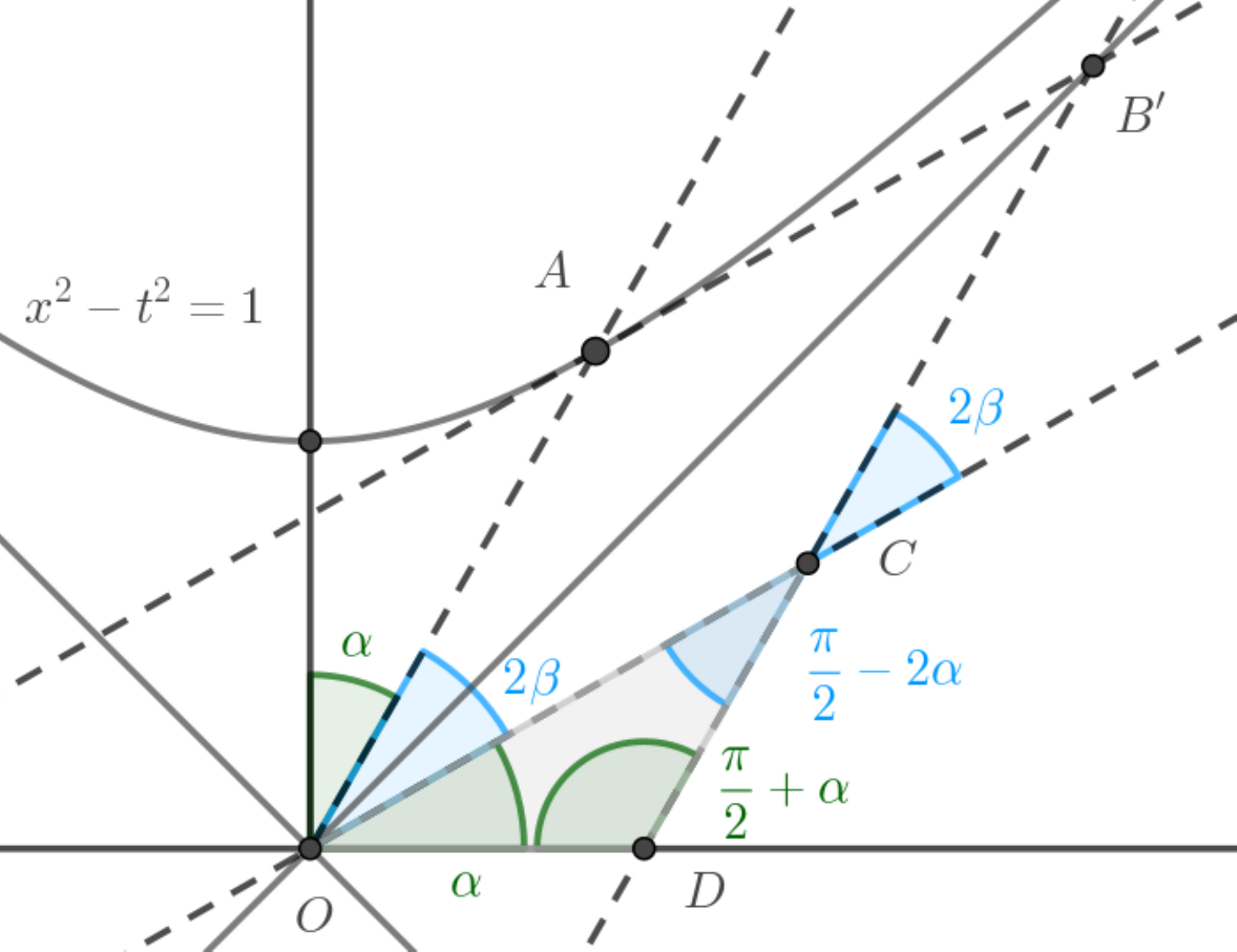}
\caption{Math.Arch.60.4, f56v; Minkowski's space-time diagram and a reconstruction of the argument.}
\label{fig:spacetime-diagram}
\end{center}
\end{figure}

\begin{figure}
\begin{center}
\includegraphics[width=0.90\linewidth]{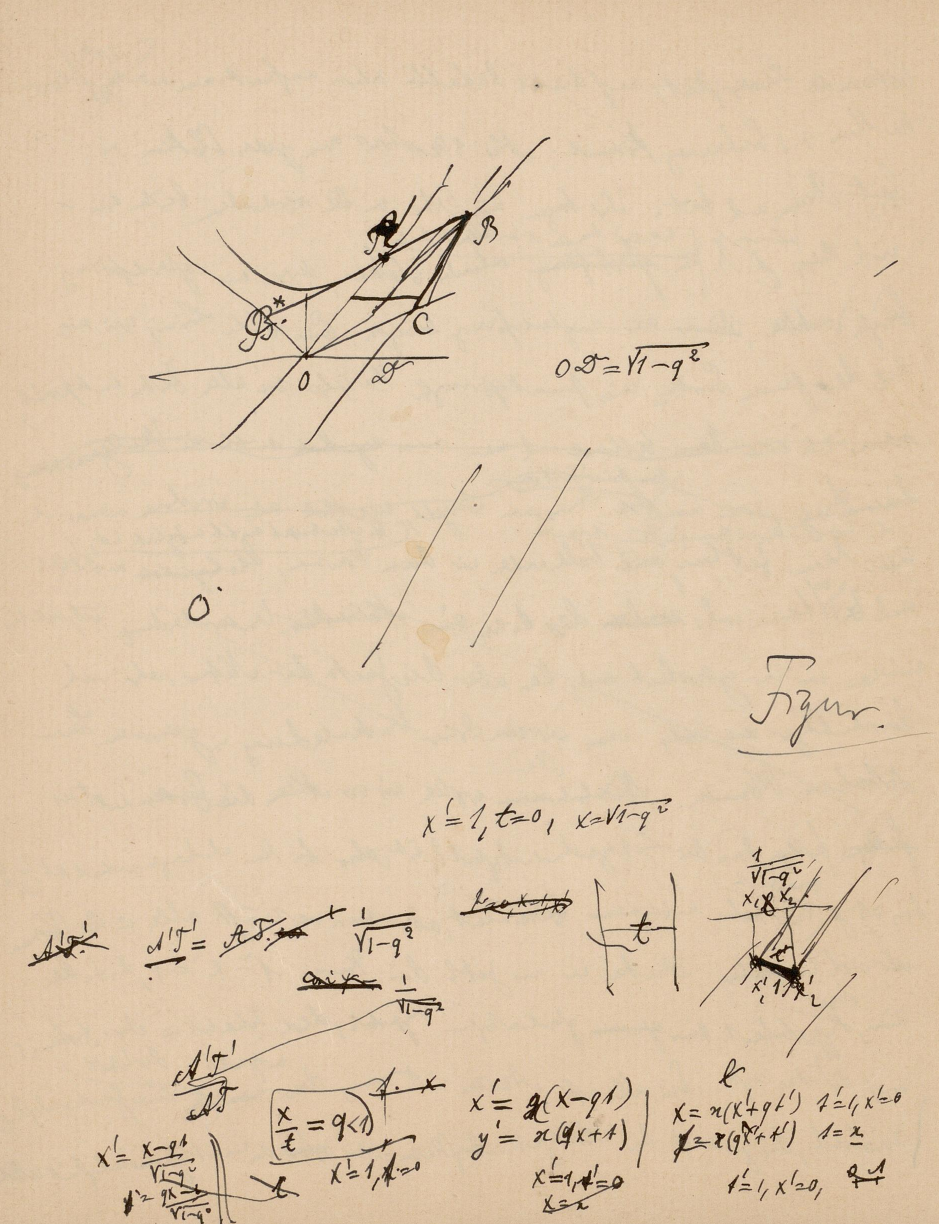}
\caption{Math.Arch.60.4,f56v; full page view. Minkowski here refers to a ``Figur'' which is not found in the manuscript.}
\label{fig:f56vfullpage}
\end{center}
\end{figure}

In Figure\,\ref{fig:spacetime-diagram}, we introduce the angle $\alpha$ between the original vertical time axis and a new, tilted one $OA$. We have $q=\tan\alpha$ denoting the
velocity of the particle moving along the tilted time axis if measured in coordinates of the original axes. By construction, $OABC$ is a parallelogram, and since $c$ has been set to 1, the asymptote $OB$ bisects the right angle between the original $x$ and $t$-axes. By construction, the new tilted axes, dashed lines in the figure, are symmetric around the asymptotes. They include an angle $2\beta$, and one can now see that in the triangle $\Delta CDO$ all angles can be expressed in terms of $\alpha$. By symmetry of the construction, the angle $\angle DOC=\alpha$. The angle $\angle DCO$ is equal to $2\beta$ since $OA$ is parallel to $CB$, and since $\beta+\alpha=\pi/4$, we have the angle $\angle OCD=\frac{\pi}{2}-2\alpha$. This leaves the third angle in the triangle $\Delta CDO$ to be $\angle CDO=\frac{\pi}{2}+\alpha$. We can now invoke the law of sines for the triangle $\Delta CDO$ to get:
\begin{equation}
\frac{OD}{OC} = \frac{\sin\angle OCD}{\sin\angle CDO} = \frac{\sin(\frac{\pi}{2}-2\alpha)}{\sin(\frac{\pi}{2}+\alpha)}
= \frac{\cos(2\alpha)}{\cos(\alpha)} = \frac{\cos^2(\alpha)(1-\tan^2(\alpha))}{\cos(\alpha)},
\end{equation}
and, setting $q=\tan(\alpha)$, we arrive at
\begin{equation}
OD = OC\cdot\cos(\alpha) \cdot (1-q^2).
\end{equation}
Obviously, this is not yet the desired contraction factor. But we also have not yet taken into account that the $x$- and $x'$-axes are not only tilted with respect to each other, but also equipped with different unit measures. A complete derivation of the contraction factor is not given in the manuscript with respect to both equations and a corresponding diagram. But a full view of f56v (see Fig.\,\ref{fig:f56vfullpage}) shows a reference by Minkowski to a certain  ``Figur''.
A figure that would correspond to this ``Figur'' is not contained in the manusccript, and my conjecture is that Minkowski produced a careful drawing for the entire argument which he then took out and reused for later expositions of the argument.

\begin{figure}
\begin{center}
\includegraphics[width=0.90\linewidth]{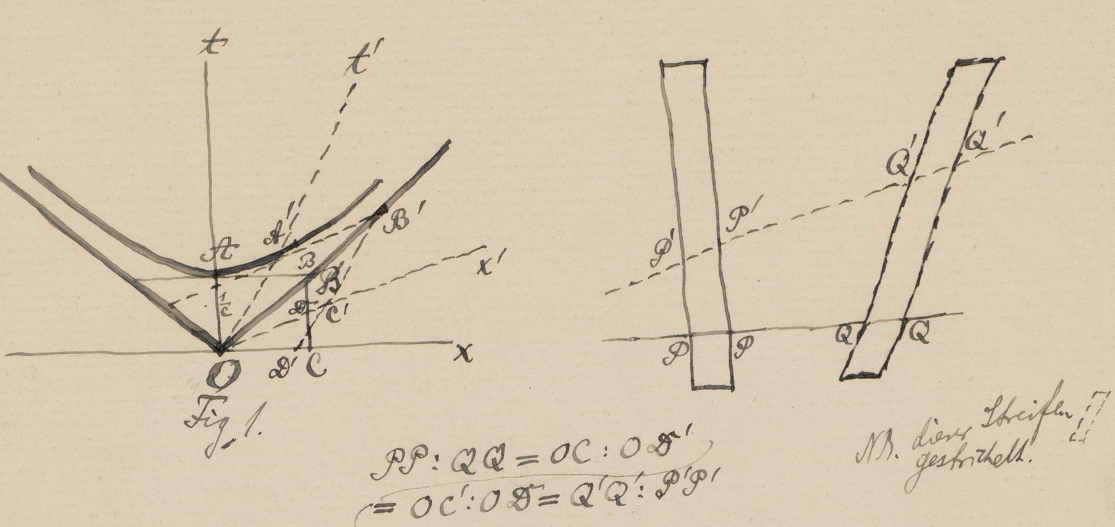}
\caption{Math.Arch.60.2.1, f25; Minkowski's figure. The drawing, without the equations, corresponds directly to the figure published in \citep[Fig.\,1]{MinkowskiH1909Raum}, except that the instruction of rendering the tilted strip with dashed lines was ignored by the typesetter.}
\label{fig:Figur-bw}
\end{center}
\end{figure}

I will therefore now continue to present the entire argument with reference to Fig.\,\ref{fig:Figur-bw}, which is close to the figure later used for the Cologne talk and to the one included in its published version. But I conjecture that this figure is already close to what Minkowski was referring to in his \emph{Ferienkurs} manuscripts. Note, however, that some of the points have been relabelled with respect to the previous diagram.

Two electrons, in uniform relative motion, $P$ and $Q$ are to be represented in a $(1+1)$-dimensional space-time diagram. Electron $P$ is at rest in the $x,t$-coordinate system, i.e.\ its space-time thread is parallel to the $t$-axis. Likewise, electron $Q$ is at rest in the $x',t'$-coordinate system, i.e.\ its space-time thread is parallel to the $t'$-axis.
The construction of the conjugate diameters and coordinate axes is done with respect to the standard hyperbola and the light-like asymptotes. The parallelogram $\parallelogram OCBA$ defines the $t$-axis and the $x$-axis, and  the units $OA$ and $OC$ on them (still for $c=1$). Likewise, the parallelogram $\parallelogram OC'B'A'$ defines the $t'$-axis and the $x'$-axis, with their corresponding units $OA'$ and $OC'$. The fact that the first parallelogram $\parallelogram OCBA$ is actually a square in Fig.\,\ref{fig:Figur-bw} depends on our choice of $c=1$ as well as on our initial arbitrary choice of orthogonal Cartesian coordinates for one of the electrons. But, as Minkowski emphasized in the second manuscript, 
\begin{quote}
[The fact] that we took the second axis at right angles, only happened for convenience, the right angle does not have any essential significance. We might as well take oblique axes.''\footnote{``Dass wir die zweite Axe rechtwinklig nehmen, geschieht nur der Bequemlichkeit, der rechte Winkel hat für uns gar keine wesentliche Bedeutung. Wir könnten auch schiefe Axen nehmen.'' (60.4, f135)}
\end{quote}
The fact that the hyperbola should be invariant implies that for the $t'$- and $x'$-coordinates of the point $A'$ we have

\begin{equation}
1 = t^2(A')-x^2(A') = OA'^2\cos^2(\alpha) - OA'^2\sin^2(\alpha). 
\end{equation}
Using $OA'=OC'$, we obtain
\begin{equation}
OC' = \frac{1}{\cos(\alpha)\sqrt{1-\tan^2(\alpha)}} = \frac{1}{\cos(\alpha)\sqrt{1-q^2}}.
\label{eq:OC1}
\end{equation}
But from the previous argument we know that for our new relabelled points ($D\rightarrow D'$, $C\rightarrow C'$), we also have
\begin{equation}
OD' = OC'\cdot\cos(\alpha) \cdot (1-q^2).
\label{eq:OC2}
\end{equation}
Inserting (\ref{eq:OC1}) into (\ref{eq:OC2}) finally gives the result
\begin{equation}
OD' = \sqrt{1-q^2},
\end{equation}
and this is the desired exact numerical expression of Lorentz contraction, if $OC=1$ is taken as unity on the $x$-axis.

\begin{figure}
\begin{center}
\includegraphics[width=0.90\linewidth]{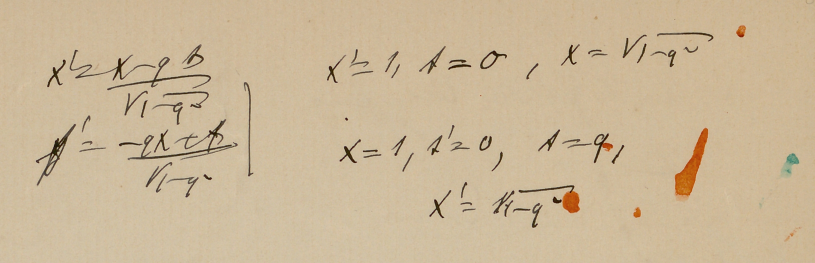}
\caption{Math.Arch.60.1.III,f34v; Minkowski on Lorentz contraction, emphasizing the symmetry of the two coordinate systems. Note also the ink stains which may indicate a direct connection with the colored transparency of Fig.\,\ref{fig:Figur-colored}.}
\label{fig:f34v}
\end{center}
\end{figure}

\paragraph{Connection with the the Lorentz transformations.}
The argument so far did not refer to the explicit form of the Lorentz transformations. It also did not follow from the sketch of Fig.\,\ref{fig:spacetime-diagram}, because the space-time diagram there did not introduce units on the axes. However, on the same page as Fig.\,\ref{fig:spacetime-diagram}, (see Fig.\,\ref{fig:f56vfullpage}), Minkowski did write down the Lorentz transformations as
\begin{align}
x' &= \frac{x-qt}{\sqrt{1-q^2}} \label{eq:Lorentz-x}\\
t'  &= \frac{qx - t}{\sqrt{1-q^2}} \label{eq:Lorentz-t}
\end{align}
and obtained the desired result of Lorentz from the first of these equation by simply setting
\begin{equation}
x' = 1, \quad t=0
\end{equation}
to obtain the Lorentz factor as
\begin{equation}
\quad x=\sqrt{1-q^2}.
\end{equation}
For a similar calculation, see Fig.\,\ref{fig:f34v}, where the same simple argument is made in a symmetric way for both primed and unprimed coordinates.

\begin{figure}
\begin{center}
\includegraphics[width=0.90\linewidth]{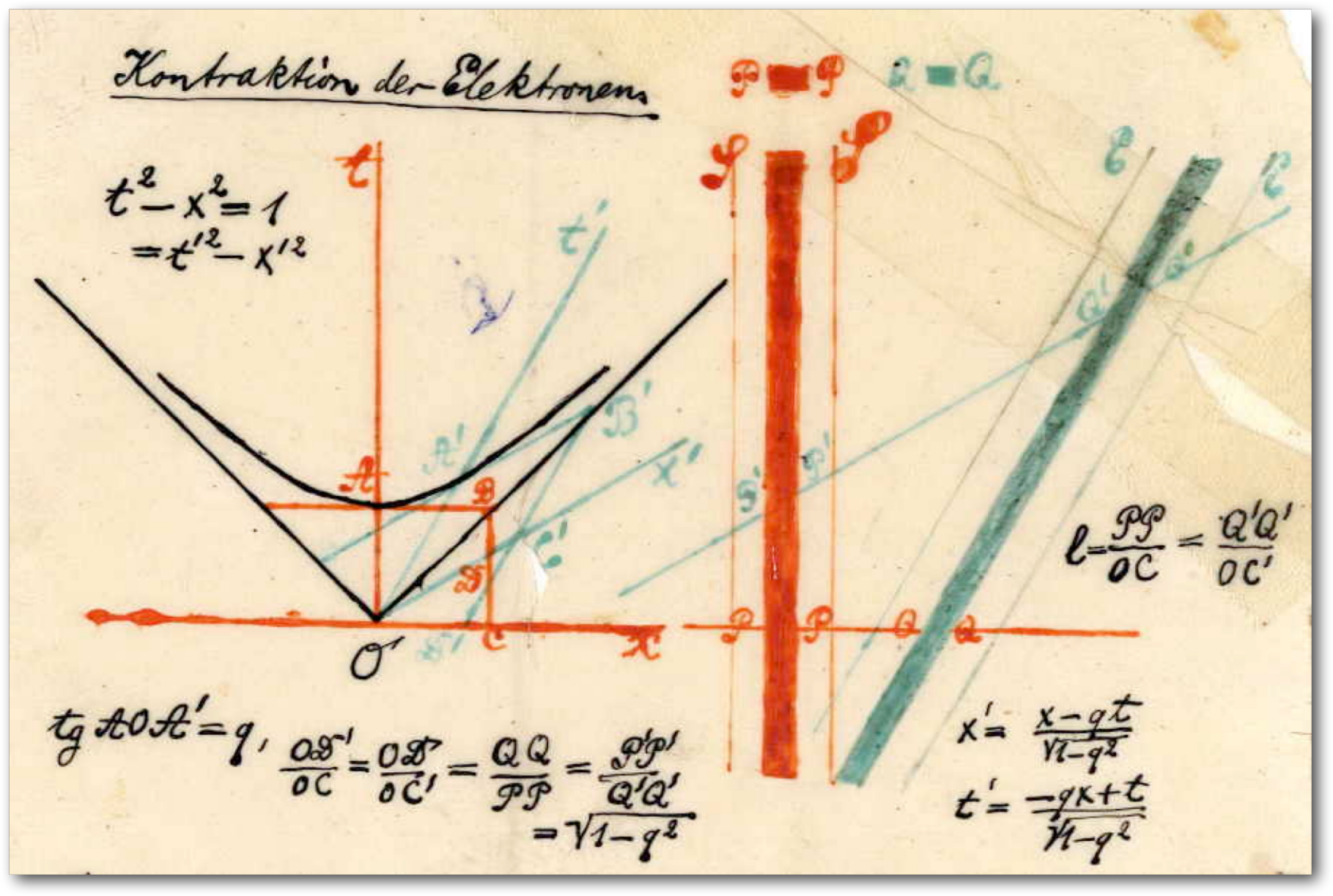}
\caption{Math.Arch.60.2.I; Minkowski's original transparency used to illustrate the geometric derivation of Lorentz contraction. Size of original: $11.5\times8.5$\,cm}
\label{fig:Figur-colored}
\end{center}
\end{figure}

The full colored transparency is shown in Fig.\,\ref{fig:Figur-colored}.\footnote{The transparency is contained in a folder 60.2.I, which otherwise contains a draft for the Cologne lecture, which is already rather close to the printed version. It has already been discussed (with reproduction) by  \cite[p.90]{GalisonP1979SpaceTime}, \citep{RoweD2009Look}, \citep[p.78]{WalterS2018Ether}, \citep[p.221]{RoweD2018Picture}.
}
It renders the $(t,x)$-coordinates and the $PP$ electron with red ink, and then $(t',x')$-coordinates and the $QQ$ electron in green ink. It also shows all relevant equations.

While it seems safe to assume that Fig.\,\ref{fig:Figur-colored} shows the transparency shown during his talk in Cologne, we can now also conjecture that the same transparency may well have been created quite a bit earlier, perhaps already for the presentation during the \emph{Ferienkurse}.  It would explain the lack of further explicit drawings of the space-time diagram of Lorentz contraction in the manuscripts and the prominent reference to some ``Figur'' on f56v (Fig.\,\ref{fig:f56vfullpage}). 

\section{Some comments on interpretation and historical context of Minkowski's notes}

We have focussed here on Minkowski's explanation of relativistic length contraction, which appears to have been the a core insight for Minkowski and which triggered the elaboration of the space-time arguments building on his \emph{Grundgleichungen} paper. It also covers roughly the content of his first manuscript. In his second manuscript for the \emph{Ferienkurs}, Minkowski also spends more time discussing further aspects. In the following, I will comment on a few points in more detail.

\subsection{The use of imaginary time}

The geometric derivation of Lorentz contraction works entirely without introducing an imaginary time coordinate. The use of an imaginary time, and specifically, the change of notation of the space time coordinates from $x$, $y$, $z$, $t$ to $x_1$, $x_2$, $x_3$, $x_4$, with $x_4=ict$ had been introduced in \citep[\S\,1]{MinkowskiH1908Grundgleichungen} and had played a prominent role in his presentation of that paper to the Academy. But 
the use of an imaginary time unit only serves heuristic purposes: it enables Minkowski to establish and exploit the analogy to the symmetry of a circle or a sphere. It may also simplify analytic trigonometric relations.

In fact, in the second manuscript for the \emph{Ferienkurs}, Minkowski added an ``instructive'' comment on the Lorentz transformations (\ref{eq:Lorentz-x}), (\ref{eq:Lorentz-t}) to the effect that ``this transformation, which we could call a space-time-distortion, can be connected with an ordinary rotation''.\footnote{``dass diese  Transformation, die wir 
als Raum-Zeit-Verzerrung bezeichnen können, mit einer gewöhnlichen Drehung in Zusammenhang gebracht werden kann'' (60.2.II, f8)}
Introducing a purely imaginary angle $i\psi$ (and using the notation lognat for the natural logarithm, tg for the tangent function),
\be
\psi = \frac{1}{2}\operatorname{lognat}\frac{1+q}{1-q},
\ee
Minkowski continued, one can write
\begin{align}
-i\operatorname{tg} i\psi &= q = \frac{e^{\psi}-e^{-\psi}}{e^{\psi}+e^{-\psi}}  \\
\cos i\psi &= \frac{1}{\sqrt{1-q^2}}, \\
\sin i\psi &= \frac{iq}{\sqrt{1-q^2}}
\end{align}
and the Lorentz transformations appear as
\begin{align}
x' &= \cos i\psi x + \sin i\psi it \\
it' &= -\sin i\psi x + \cos i\psi it
\end{align}
in complete analogy to ordinary rotations in euclidean space.

\subsection{The perception and criticism of Einstein's work}

On the first day of his \emph{Ferienkurs}, Minkowski gave an explicit discussion of the Michelson-Morley interference experiment, as a motivation for the Lorentz contraction hypothesis. It appears, in general, that Minkowski in the second manuscript addressed himself more to an audience of physicists. He refers to Lorentz's work, of course, but also to Einstein's, and it is these two theorists only that he introduced to his auditors.  It is interesting to see Minkowski struggle to find the best characterization of Einstein's work, both in comparison to Lorentz's as well as to his own approach. 

In the beginning, he mentioned Lorentz's contraction hypothesis, which he called ``well-known,'' or as he corrected himself, as ``perhaps known'' to his auditors. After a brief characterization of the hypothesis, he continued to introduce ``the young physicist A. Einstein in Bern'', who had worked through the facts contained in Lorentz's hypothesis with a more liberated frame of mind (``in einer freieren Auffassung''). The characterization of ``more liberated'' he then corrected, first, to ``highly significant'' (``höchst wichtigen''), before settling for a simple ``new'' (``neuen''). In heavily redacted phrasing, he then suggested that Einstein had analyzed Lorentz's notion of a local time (``Ortszeit'') from an axiomatic point of view and concluded that Lorentz's postulate was not to be seen as a concrete hypothesis, but rather as a ``new'' conception, or changed imagination, of the concept of time.  

Following this positive characterization, he strives to find a good phrase which would capture the difference of his own approach. He suggested that Einstein would still be ``swimming in the wake of Lorentz'' (``noch ganz im Fahrwasser von Lorentz schwimmend''). He suggested that Einstein would ``at some places'' get ``entangled in contradictions'' (``verwickelt er sich in Widersprüche'') but would continue along the direction that he had taken (``in seiner eingeschlagenen Bahn weiter arbeitet''), as witnessed by the recent publication of the \emph{Jahrbuch} paper. He sees Einstein as having ``made up'' (``zurechtgelegt'') a certain visualization and an interpretation using clocks and signals. Unfortunately, Einstein's interpretation had already been taken up by other authors: 
\begin{quote}
... such that soon a hopeless atmosphere of signal works has gotten the upper hand.\footnote{%
``sodass bald eine heillose Signalwerkluft eingerissen ist.'' (f51/M02)}
\end{quote}
In the margin, he noted:
\begin{quote}
In my opinion, these interpretations can only obscure the true facts.\footnote{%
``Diese Interpretationen sind nach meiner Meinung nur geeignet den wahren Sachverhalt zu verdunkeln.'' (f51/M03)}
\end{quote}
He then stated his impression that Einstein had not yet achieved ``full understanding of the facts''(``volle Einsicht der gegebenen Tatsachen'').

What follows is an interesting statement about his own instruction of Einstein as a student at the Polytechnic and the latter's mathematical knowledge and skills. This passage has already been published by Pyenson (\citeyear[p.\,72, n\,7]{PyensonL1977Minkowski}). It seems worthwhile to give a slightly emendated translation based on a fully diplomatic transcription here:
\begin{quote}
I believe that this is essentially due to the limitations of his mathematical tools, and I am the one who can say so, because Einstein received his mathematical training from me. Although he was one of the best at the time, the mathematical knowledge he was able to take with him from the Polytechnic in Zurich, where so many other things were going on, was of course imperfect. I mention this here so that you will at least grant me some authority in my judgment of Einstein, even if I do not know how much authority you will otherwise grant me with regard to the correctness of the judgments in physical matters that I now want to make.  To say my own opinion at once, that Lorentz's hypothesis, correctly understood, is a law of nature of the first rank, I would even say the first of all laws of nature, on the one hand because it deals with the most basic concepts of all scientific knowledge of nature, with the conception of space and time, and on the other hand because of the quite extraordinary consequences of this law, the most surprising of which have not yet been noticed.\footnote{%
``Es liegt das wesentlich \del{daran, dass um dieses} \intd{\del{solches}} \del{zu erreichen, eine} glaube ich, an \del{der} \intd{einer} Beschränktheit seiner mathematischen Hilfsmittel und ich \del{darf} \intd{bin derjenige, der} dieses wohl behaupten darf, denn Einstein hat \del{seinerzeit} seine mathematische Ausbildung durch mich bezogen, obwohl er \intd{seinerzeit} der besten einer war, so sind doch die mathematischen Kenntnisse, die er vom Polytechnickum in Zürich mitnehmen konnte, wo so viele andere Dinge voranstanden, natürlich unvollkommen gewesen. Ich \del{sage} \intd{erwähne} dieses \intd{hier deshalb}, damit sie mir wenigstens eine Autorität in meinem Urteil über Einstein zuerkennen, wenn ich auch nicht weiss, wieviel Autoritäten sie mir \intd{sonst} in Bezug auf die Richtigkeit der Urteile in physikalischen Dingen, die ich nun abgeben will, einräumen wollen. Nämlich \intd{gleich} meine eigene Meinung zu sagen, so bedeutet jene \intd{ist meine Überzeugung dass?} \intd{Lorentzsche} Hypothese richtig aufgefasst, ein Naturgesetz allerersten Ranges, ich möchte sogar geradezu sagen, das erste aller Naturgesetze, nämlich \del{nicht bloss} \intd{einerseits} deshalb, weil es sich um die ursprünglichsten Begriffe aller \del{Forschung} \intd{\del{naturwissenschaftlich} Naturerkenntnis um die Auffassung von} Raum und Zeit handelt \del{sondern} \intd{dann \del{aber}} wegen \del{der} ganz ausserordentlichen Konsequenzen dieses Gesetzes, \del{über die} \intd{\del{bisher} \del{in dieser Form} von denen die überraschendsten bisher} noch gar nicht bemerkt worden sind.'' (f52/M03)}
\end{quote}

Minkowski's discussion of Lorentz and Einstein as point of departure for his own presentation ends here. But at a later point he came back to an assessment of Einstein in the following passage which takes up his earlier concerns:
\begin{quote}
Einstein has come up with an interpretation with light signals, resting and moving clocks, which has already been propagated in the literature, which is based on Lorentz's explanation of the concept of local time, which, as it seems to me, still strongly obscures the innermost core of the matter. This seems to me to be particularly clear from the way in which Einstein, in his latest article, continues to generalize his conceptualizations to coordinate systems that relate to others in accelerated motion.\footnote{%
Einstein hat sich eine Interpretation mit LichtSignalen, ruhenden und bewegten Uhren  zurechtgelegt, die sich auch schon weiterhin in der Litteratur fortgepflanzt hat, die sich an der Lorentzschen Erläuterung des Begriffs der Ortszeit anlehnt, die  wie mir scheint, doch den innersten Kern der Sache noch stark verdunkelt. Das scheint mir besonders klar hervorzugehen aus der Art, wie nun Einstein in seinem letzten Aufsatze weiter zu Verallgemeinerungen seiner Begriffsbildungen auf Koordinatensysteme, die sich  in beschleunigter Bewegung gegen andere beziehen, kommen will. (f62/M13)}
\end{quote}
This entire paragraph was later struck through as well. It seems that Minkowski really struggled to define his own approach in contrast to Einstein's. The reworking of these passages presumably also reflects Minkowski's meeting with Lorentz in Rome. Elsewhere, in a later manuscript, he wrote:
\begin{quote}
Lorentz himself told me in a conversation in Rome that Einstein's merit is to have recognized that the time of one electron is just as good as that of the other, i.e. to have recognized the equivalence of $t$ and $t'$.\footnote{%
``Lorentz selbst sagte mir gesprächsweise in Rom, erkannt zu haben, dass die Zeit
des einen Elektrons ebensogut wie die des anderen ist, d.h. die Gleichwertigkeit von $t$ und
$t'$ erkannt zu haben, das Verdienst von Einstein ist.'' (60.4, f11/M12) This passage is also cited by \cite[p.\,67]{WalterS1999Minkowski}.}
\end{quote}
The passage immediately following here was deleted and rephrased repeatedly.\footnote{The passage reads in diplomatic transcription: ``\del{So wird} \del{So wird der Begriff} Damit ist zunächst \del{der Begriff} die Zeit als eine durch die Erscheinungen eindeutig \del{fixierte} festgelegter Begriff \del{ausgeschaltet} \intd{abgesetzt}. \del{Auch den} \del{Das Gleiche} Den Begriff des Raumes \del{hatten wohl ebenfalls in analoger Weise} entsprechend zu traktieren \del{modifizieren} ist wohl \intd{nur?} als \del{ein}
Erzeugnis \intd{rein?} mathematischer Kultur \del{zu} einzutaxieren.'' (60.4, f11/M12)}
It seems that Minkowski interpreted the progress of his own advancement to consist in fully relativizing also the spatial dimension. In a marginal comment, he added explicitly: 
\begin{quote}
Neither Lorentz nor Einstein shook up the concept of space.\footnote{``Am Begriff des Raumes rüttelten weder Lorentz noch Einstein.'' (60.4, f10v)}
\end{quote}
And he suggested that it is only ``mathematical culture'' which would lead to treating the spatial dimension in the same way as the temporal dimension and include it in a full relativization of space-time. 

With respect to the new space-time diagram representation, Minkowski's new insight would refer specifically to the identification of the tilted $x'$-axis as the proper spatial dimension associated with an electron whose world line is given by the $t'$-axis. After all, the entire space-time diagram representation with the standard hyperbola had been introduced by Minkowski. Implying that Einstein had not taken the concept of space into question seems, however, somewhat prejudiced. It is true, however, that Einstein was initially rather critical of Minkowski's approach which he found mathematically overambitious\footnote{For further discussion of this point, see \cite{WalterS1999Minkowski}.} and, in particular, criticized a sweeping disregard of the difference between space and time, as he explained, e.g., in a letter to Arnold Sommerfeld of July 1910:
\begin{quote}
The conditions for events (differential equations) are symmetrical in four dimensions; this realization makes finding those conditions easier. The limit of the significance of the four-dimensional approach seems to me to consist in the fact that in the solutions of those equations that are of interest to us the four dimensions do not appear in the same manner. \citep[Doc.\,211]{CPAE05}
\end{quote}

\subsection{The relationship between mechanics and electrodynamics}

In his first manuscript, Minkowski had set the speed of light equal to unity, $c=1$, almost throughout. In his second manuscript, he included this factor as a variable parameter to discuss the relationship between the covariance group of Maxwellian electrodynamics, which he calls $G_c$, and the covariance group of the equations of classical mechanics, expressing Galilean invariance. The latter he identified as a limiting case for infinitely large $c$, and called it $G_{\infty}$. Here he explicitly discussed the geometric implications of the limiting process of taking very large values for the speed of light: ``The hyperbola flattens out.'' (``Die Hyperbel flacht sich ab.'' 60.2.II, f8). After discussing Newtonian mechanics as a limiting case, he commented:
\begin{quote}
Much has been said about the fact that the laws of electrodynamics are in contradiction to those of Newtonian mechanics, that [the principle of] action and counteraction no longer holds in electrodynamics and so on. In my opinion, the contradiction will be completely resolved in such a way that Newtonian mechanics only represents an approximation to reality. An exact mechanics would have to be put in its place, which would not differ at all from Newtonian mechanics in the observed phenomena, but which, nonetheless, will lead to other consequences in connections with electrodynamics.\footnote{%
``Es ist viel davon geredet worden, dass die Gesetze der Elektrodynamik denen der Newtonschen widersprechen, [das Prinzip von] Wirkung und Gegenwirkung in der Elektrodynamik aufhört u.s.w. Nach meiner Ansicht würde der Widerspruch sich völlig heben in der Weise, dass eben die Newtonsche Mechanik nur erst eine Approximation an die Wirklichkeit vorstellte, und \del{über} an ihre Stelle eine exakte Mechanik zu setzen wäre, welche in den beobachteten Erscheinungen \del{an die kaum von} gar keine Differenz gegen die Newtonsche Mechanik ergibt, welche aber, doch in den Zusammenhängen mit der Elektrodynamik zu anderen Konsequenzen hinführt.'' (60.4, f21/M07)}
\end{quote}
In fact, the second manuscript engages in an explicit discussion of the necessary adaptation of fundamental concepts of mechanics, like force and energy, the equations of motion, and the mass-energy equivalence. In a passing reference to contemporary debates about an electromagnetic world picture, Minkowski concluded with a final remark that he would liked to have expanded on that topic but did not have enough time to do so:
\begin{quote}
Gentlemen, What I have been able to suggest to you in such a short time is, of course, very sketchy. I would have liked to discuss attempts to explain the concept of mass in purely electromagnetic terms, but I was forced to limit myself to the subject matter.\footnote{%
``M.H. Was ich Ihnen in der kurzen Frist andeuten konnte, trägt naturgemäss ei	nen sehr lückenhaften Charakter. Ich hätte noch gern von den Versuchen, den Begriff der Masse rein elektromagnetisch zu erklären gesprochen, musste mich aber in meinem Stoffe notwendig einschränken.'' (60.4, f18/M12)}
\end{quote}
Minkowski embraced the changes in world view suggested by the new relativistic physics:
\begin{quote}
The more I work on these questions myself, the more I am convinced that the world is no longer as we once thought it to be; that space geometry exists, one could say, only as a chapter of mechanics.\footnote{%
``Je mehr ich selbst an den Fragen weiter arbeite, befestigt sich in mir umsomehr der Eindruck, dass in der Tat die Welt nicht mehr so ist, wie \del{ich} wir sie uns sonst dachten, eine \intd{Raum}Geometrie gibt es \del{eigentlich} \del{nur mehr}, so kann man sagen, nur noch als Kapitel der Mechanik.'' (f18/M12)}
\end{quote}

\subsection{Political metaphors}

The charm of the manuscript notes is that Minkowski not only used colorful metaphors and tried out unusual phrases, but he also reflected explicitly on that use of language. Thus, he continued his final remarks to his \emph{Ferienkurs} audience:
\begin{quote}
In my publications, I have endeavored to use more diplomatic language so as not to provoke too much contradiction. Here before you, I have stuck more to the saying, which is not always pedagogical, but which in a circle of only pedagogues can always hold true: He who knows the truth and does not tell it... I thank you sincerely for the interest with which you have followed me, and I am happy to respond to any further inquiries.\footnote{%
``In meinen Publikationen habe ich mich einer mehr diplomatischen Ausdrucksweise befleissigt, um nicht zu sehr Widerspruch herauszufordern. Hier vor Ihnen habe ich mehr mich an den Spruch gehalten, der ja nicht immer pädagogisch ist, den Sie aber in einem  Kreise von \intd{nur} Pädagogen \del{immer} \intd{allezeit noch} gelten darf: Wer die Wahrheit kennt und sagt sie nicht. Ich danke Ihnen herzlich für die Teilnahme, mit der Sie mir gefolgt sind, und bin noch gern bereit auf Anfragen zu erwidern.'' (f18/M12)}
\end{quote}
The elliptic reference to the ``saying'' alludes to a rhyme of the poet, journalist and founding member of one of the first German fraternities, August Daniel von Binzer (1793--1868). In his very popular fraternity song ``Toast!'' (``Stosst an!'') of 1817, one of the verses runs
\begin{quote}
Toast! Long live the free word! Hurrah! \\
Whoever knows the truth and doesn't speak it,\\ 
will truly remain a miserable wretch. \\
The lad is free!\footnote{%
``Stoßt an! Freies Wort lebe! Hurra hoch! //
Wer die Wahrheit kennet und saget sie nicht, // 
der bleibt fürwahr ein erbärmlicher Wicht. //
Frei ist der Bursch!''}
\end{quote}
Another curious example of Minkowski's undiplomatic language in his manuscripts is the following quote. After emphasizing the arbitrariness of taking the $x,t$-lines or the $x',t'$-lines as coordinate axes, Minkowski quips
\begin{quote}
It is therefore purely a matter of standpoint as to what is at rest and what is in motion. Thus, from a modern perspective, Galileo is again wrong with his Eppur si muove, and I hope that this remark will make modernism very popular in Vatican circles.\footnote{%
``Es ist also reine Auffassungssache, was ruhend, was bewegt ist. So hat Galilei mit seinem Eppur si muove nach modernem Standpunkte wieder Unrecht und ich hoffe durch diese Bemerkung den Modernismus in vatikanischen Kreisen sehr beliebt zu machen.'' (f64/M15)}
\end{quote}
Alluding to the conflict between Galilei and the Catholic Inquisition, the Jewish mathematician will probably have provoked a good laugh with this joke in his audience of physicists and mathematicians in predominantly Protestant Göttingen.

Finally, I want to discuss another example of careful rephrasing by Minkowski. This passage is not part of the manuscripts from late March or early April 1908, which we have discussed in this paper but from a later draft manuscript of the Cologne talk. The passage was already printed as a black-and-white facsimile by Galison (\citeyear[Fig.\,3 on p.\,99]{GalisonP1979SpaceTime}) but can only deciphered from the original itself or a  high-quality color scan of it. A diplomatic transcription will be given in a footnote,%
\footnote{%
``M.H. Die Anschauungen über Raum und Zeit, die ich Ihnen heute entwickeln will, sind \del{entstanden} erwachsen auf \del{rein} \intd{durchaus} experimentell-physikalischem Boden. 

\del{Sie} \del{Darauf} \del{Darauf} \intd{Darin liegt} \del{beruht} ihre Stärke. 

Ihr Charakter aber ist \del{derart} \del{höchst}\intd{\del{derart} höchst \del{derart} \del{äusserst}} \intd{gewalttätig} revolutionär, \intd{derart, dass wenn wir sie in 100 Jahren durchdringen \intd{\del{dermassen}}, woran ich glaube \del{werden}} \del{Raum und Zeit werden jeder für sich völlig depossediert und nur dass} es \del{fast unnütz} verpönt \del{ist} \intd{sein wird} \intd{noch} davon zu sprechen, wie wir \del{uns} \intd{bislang uns Mühe gaben} Raum und Zeit \del{bisher vorzustellen} \intd{\del{bislang}} \intd{zu verstehen} \del{trachteten} \intd{\del{Mühe gaben}}.

\intd{\del{Denn} Denn} Von Stund an \intd{\del{nämlich}} \del{sollen} \intd{\del{werden} sollen} Raum für sich und Zeit für sich \intd{zu} \intd{\del{als getrennte Begriffe}\intd{\del{Wesen}}\del{Existenzen}} völligen \del{depossediert} \intd{\del{aufgehoben} \del{depossediert} \del{sein} Schatten herabsinken,} \del{und} \intd{\del{wird} \del{soll} wird} nur noch ein durch Verschmelzung beider \del{gewonnener} \intd{der Beiden} \del{gewonnener} \intd{\del{geschaffener} gewonnener} Begriff, der \intd{der der provisorischen Erinnerung} bis zur \del{Erfindung} \intd{Erinnerung} eines \del{prägnanteren} eigenartigen \del{Namens} \intd{den Namen \del{---?meinetwegen}} Welt heissen \intd{geben will} \intd{\del{genannt} geheissen, heissen mag} eine freie Existenz \del{haben} zeigen.'' (60.2, f1/M01)
}
here I want to highlight the following points.

In this passage, we see Minkowski working on his first introductory paragraph for the Cologne lecture, the famous opening paragraph quoted in the beginning of this paper. While in the printed version he said about the new view of space and time that ``their tendency is radical'', he here tried out different expressions whose political connotation was much more manifest. Thus, he first used the word ``violent'' (``gewalttätig'') and then considered the word ``revolutionary'' (``revolutionär''). He anticipated that after a long period, possibly of a hundred years, we will finally fully have understood the new conception, and then it will be ``useless'' (``nutzlos'') or even ``frowned upon'' (``verpönt'') to consider space and time as separate entities. And then Minkowski tried out another metaphor from the political realm. He wrote that space and time each  would be ``depossessed'' (``depossediert''), a term which he introduced and deleted again three times in the draft paragraph. The word ``depossediert'' is no longer used or even known in today's German but was a well-known term at the time. In a Conversation Lexicon of 1906, the word is explained as ``put out of possession, dethrone'' (``aus dem Besitz setzen entthronen'')\footnote{Meyers Gro{\ss}es Konversations-Lexikon, Band 4, Leipzig 1906, pp.\,648--649.}, and the entry continues to explain that the term ``depossessed'' was used for monarchs that had been dismissed from government in Italy in 1859 and 1861 and in Germany in 1866. In general, the term may refer to loss of wealth, privilege, or power, and the special role of depossessed aristocratic families in Prussia were subject of a special article 57 in the Introductory Act to the Civil Code of 1896 (``Einführungsgesetz'' for the ``Bürgerliches Gesetzbuch''). Similar to his anti-Catholic quip, Minkowski's metaphorical use of the word in the context of a new concept of space and time would have thus invoked a modernist, republican, anti-aristocratic stance.

\section{Concluding remarks}

The analysis of Minkowski's unpublished, private notes on space and time from 1908 revealed interesting insights in several respects. As regards his intellectual development, it reflects a discovery process. The first manuscript about the geometric explanation of Lorentz contraction of electrons seems to document Minkowski's discovery of the possibility of a mathematical derivation and explanation of a significant physical hypothesis. Even if we don't see him discovering the argument itself for the first time in these notes, we can still see how he tried to expound it and make it accessible to an audience of physicists and mathematicians. For Minkowski, the successful derivation of the contraction factor of special relativistic length contraction was a major triumph of the application of sophisticated mathematical tools, methods and concepts in the theoretical natural sciences. It was a victory in a general program of mathematizing science. As such it was in full agreement with Hilbert's program of an axiomatization of physics as postulated in his 6th mathematical problem of 1900. The fact that Minkowski elaborated his argument for the model of an extended electron, rather than as generic relativistic length contraction, reflects the historical context of early relativistic electrodynamics. Minkowski's frequent emendations, rephrasings, reveal him not only as a witty author striving for a careful phrasing of his words. They also illustrate the self-reflection of an emerging independent discipline of mathematical physics.

\section*{Acknowledgments}

Preliminary versions of this paper were presented at the workshop ``Geometry and Society'' in Brno, Czechia, June 13, at the Minkowski conference in Kaunas, Lithuania, June 20, 2024, at the RheWeSe, January 31, 2025, in Mainz, Germany, at the ITEM seminar of the ENS in Paris, May 27, 2025, and in Bonn-Königswinter, May 29, 2025. 
I wish to thank Scott Walter for inspiring discussions and helpful comments, and Emmylou Haffner and the Institut des textes et manuscripts modernes (ITEM) at the ENS, Paris, for their hospitality.

\bibliographystyle{apalike}
\bibliography{minkowski}

\end{document}